\documentclass[a4paper,11pt]{article}
\usepackage{jheppub}
\newcommand{\gev}{\ensuremath{\,{\mathrm{GeV}}}}
\newcommand{\mev}{\ensuremath{\,{\mathrm{MeV}}}}

\newcommand{\xip}{\ensuremath{\xi^\prime}}
\newcommand{\alphap}{\ensuremath{\alpha^\prime}}
\newcommand{\betap}{\ensuremath{\beta^\prime}}
\newcommand{\xipp}{\ensuremath{\xi^{\prime\prime}}}
\newcommand{\etapp}{\ensuremath{\eta^{\prime\prime}}}

\newcommand{\ee}{\ensuremath{e^+ e^-}}
\newcommand{\tat}{\ensuremath{\tau^+ \tau^-}}
\newcommand{\tlnn}{\ensuremath{\tau^- \to \ell^- \bar{\nu}_\ell\nu_\tau}}
\newcommand{\tmnn}{\ensuremath{\tau^- \to \mu^- \bar{\nu}_\mu\nu_\tau}}
\newcommand{\tmnnf}{\ensuremath{\tau\to \mu\nu\nu}}
\newcommand{\atmnn}{\ensuremath{\tau^+ \to \mu^+ \nu_\mu \bar{\nu}_\tau}}
\newcommand{\menn}{\ensuremath{\mu^- \to e^- \bar{\nu}_e\nu_\mu}}
\newcommand{\mennf}{\ensuremath{\mu\to e\nu\nu}}

\title{\boldmath A new method for the measurement of the Michel parameters that describe the daughter muon polarization in the $\tau^- \to \mu^- \bar{\nu}_\mu\nu_\tau$ decay}

\author[a,b]{D. Bodrov,}
\author[b,a]{P. Pakhlov}
\affiliation[a]{International Laboratory of Elementary Particle Physics, National Research University Higher School of Economics, 20 Myasnitskaya Street, Moscow 101000, Russia}
\affiliation[b]{Laboratory of Heavy Quarks and Leptons, P.N. Lebedev Physical Institute of RAS, 53 Leninskiy Prospekt, Moscow 119991, Russia}
\emailAdd{dbodrov@hse.ru}
\emailAdd{ppakhlov@hse.ru}
\date{\today}

\abstract{This paper provides a detailed description of the method for the first direct measurement of all Michel parameters in the $\tmnn$ decay related to the polarization of the daughter muon. An application of the suggested method in the existing and future experiments at \ee\ colliders is considered. We have performed a feasibility study for the future Super Charm-Tau Factory and Belle~II experiments. For the first one, the sensitivity to the Michel parameters $\xip$, $\xipp$, $\etapp$, $\alphap/A$, and $\betap/A$ is estimated. For the latter, only one Michel parameter, $\xip$, for which the sensitivity is maximum, is considered.}

\arxivnumber{}

\begin{document}
\maketitle
\flushbottom

\date{\today}

\section{Introduction\label{sec:intro}}

The Standard Model (SM) has proven its validity in describing particle physics in multiple experimental tests, at least within the energy ranges and measurement accuracy available to experiment. However, due to the incompleteness (dark energy, dark matter, baryon asymmetry, etc) and unnaturalness (hierarchy problem, fine-tuning, strong CP problem, etc) of the SM, it is to be expected that sooner or later, experiments might report disagreement with the SM predictions, hence revealing new physics (NP).

The SM postulates maximum parity violation in weak decays via charged current since only left-handed fermions participate in vector interaction. This postulate is formulated to ensure consistency of the theory to experiments, mainly the precise measurements in the muon decay~\cite{Fetscher:1986uj, Langacker:1988fp, Zyla:2020zbs}. Deviations from this behavior in other processes can be caused by changes in the couplings of $W$-boson with fermions or by interactions mediated by new gauge or charged Higgs bosons~\cite{Bryman:2021teu}. Therefore, the conjecture of the $V-A$ Lorentz structure of the charged currents needs to be verified preferably in all possible processes caused by them. Indeed, in many SM extensions, an admixture of right-handed currents is predicted non-universal for various processes~\cite{Herczeg:1985cx}.

In $\tau$-decays, due to the much larger mass of the $\tau$ compared to the muon, the relative contribution of NP processes with an ``incorrect'' Lorentz structure can be enhanced~\cite{Krawczyk:2004na,Chun:2016hzs}. Pure leptonic $\tau$-decays provide a clean laboratory to test the structure of the weak currents since there are no QCD-associated uncertainties in their calculations.

The interaction of charged currents in $\tau$-decays in a general form that allows for a non-SM contribution but does not violate Lorentz invariance can be expressed in terms of so-called Michel parameters~\cite{Michel:1949qe}. At present, four Michel parameters have been measured with high accuracy in $\tau$-decays~\cite{Zyla:2020zbs}, and the obtained values are in agreement with the SM. They completely describe the differential decay width, integrated over the neutrino momenta and summed over the daughter lepton spin. The measurement of the remaining Michel parameters requires knowledge of the daughter lepton polarization, and it has not been done yet. The exception is two parameters obtained in radiative leptonic $\tau$-decays by the Belle collaboration~\cite{Shimizu:2017dpq}. However, this measurement still suffers from large uncertainties that limits the sensitivity to NP parameters. It is also possible to measure the Michel parameters sensitive to the polarization of the daughter lepton in five-body leptonic $\tau$-decays~\cite{Flores-Tlalpa:2015vga}. Such an approach was tested by the Belle collaboration, though only an estimation of the method sensitivity has been presented yet~\cite{Sasaki:2017unu, Sasaki:2017msf}.

Thus, for the complimentary test of the theory, it is necessary to find a method that gives access to the daughter lepton polarization. The development of super flavor factories such as Belle~II~\cite{Kou:2018nap} and Super Charm-Tau Factory (SCTF\footnote{There are two similar projects called Super Charm-Tau Factory and Super Tau-Charm Factory. For both factories, we use the notation Super Charm-Tau Factory or SCTF.})~\cite{Bondar:2013cja, Luo:2018njj}, thanks to their extremely high luminosity and correspondingly enormous data statistics, opens up a possibility to apply new methods based on the study of rare processes that were not previously considered due to their practical inapplicability with the modest statistics of the past experiments. The expected integrated luminosity of Belle~II is $50\,\text{ab}^{-1}$, which corresponds to $4.6\cdot 10^{10}$ $\tau^+\tau^-$-pairs, and of SCTF is $10\,\text{ab}^{-1}$, which corresponds to $2.1\cdot 10^{10}$ $\tau^+\tau^-$-pairs. In this work, we discuss a method based on a rare process of decay-in-flight of the daughter muon from the $\tmnn$\footnote{Charge conjugation is implied unless otherwise is indicated.} decay in the tracking detector, which opens opportunities to measure the missing Michel parameters.

The idea to use muon decay to access its polarization in the $\tmnn$ decay was suggested in~\cite{Fetscher:1990su}, where it was proposed to use stopped muons. Due to the $P$-violation in the $\menn$ decay, the electron emission angle correlates with the muon spin, providing information on its direction. Such a method with stopped muons suffers from a practical disadvantage: it requires a specific sophisticated detector that should stop energetic ($\sim 1\gev$) $\mu^+$ while not depolarizing them ($\mu^-$ will be depolarized anyway~\cite{Mann:1961zz}) and detect $\sim 50\mev$ daughter electrons. Installation of such a detector is impracticable in modern experiments. 

Recently, it was proposed to use muon decay-in-flight in the tracking system of the detector to measure the muon polarization from $\tau$ to obtain the Michel parameter $\xip$ in the future experiment at the Super Charm-Tau Factory~\cite{Bodrov:2021vkn, Bodrov:2021hfe}. For such a measurement in the experiment, it is necessary to reconstruct tracks of both the muon from the $\tau$-decay and the secondary electron from the muon decay. With a typical radius of tracking detectors in modern \ee\ experiments $\sim 1$~m, the probability of muon to decay-in-flight in the tracker is extremely small ($\sim 10^{-4}-10^{-3}$). Nevertheless, the specific signature of such events (track kink) makes it possible to select signal events without a large background. The main background, hadrons ($\pi^\pm,~K^\pm$) decaying in-flight, are mainly two-body decays, leading to a narrow line in the spectrum of the daughter particle in the parent's rest frame. Such physical background can be effectively vetoed if a kinks reconstruction algorithm is implemented in the track reconstruction~\cite{Bodrov:2021vkn, Bodrov:2021hfe}. In this paper, we demonstrate that the large statistics of the experiments at the SCTF and Belle~II will allow to achieve a good accuracy of measurement for all Michel parameters in the near future. 

The proposed method is charge-universal compared to those that utilize stopped muons, since in the latter, $\mu^-$ are depolarized in the matter after stopping. Usage of decay-in-flight muons allows measuring the Michel parameters with the same accuracy in both $\tmnn$ and $\atmnn$ decays. Thus, one can separately analyze both charges to test CPT invariance or merge two samples under the CPT invariance assumption to increase precision.

\section{Method\label{sec:theory}}

Let us recall the general formalism for describing the $\tmnn$ decay of our interest. The most general form of the Lorentz invariant, local, derivative-free, lepton-number conservative Hamiltonian of the four-fermion interaction~\cite{Michel:1949qe} leads to the following matrix element of the $\tlnn$ decay ($\ell=e$ or $\mu$), written in the form of helicity projections~\cite{Scheck:1984md,Mursula:1984zb,Fetscher:1986uj}:
\begin{eqnarray} \label{eq:1}
M=\dfrac{4G_F}{\sqrt{2}}\sum_{\substack{\rho=S,~V,~T\\
\varepsilon,~\mu=R,~L} }g^\rho_{\varepsilon\mu}\left\langle \bar {\ell}_\varepsilon \left| \Gamma^\rho \right| (\nu_\ell)_\alpha \right\rangle
\left\langle (\bar \nu_\tau)_\beta \left|
\Gamma_\rho \right| \tau_\mu \right\rangle \,, \nonumber\\
 \Gamma^S=1, \quad\Gamma^V=\gamma^\mu, \quad \Gamma^T=\dfrac{1}{\sqrt{2}}\sigma^{\mu\nu}=\dfrac{i}{2\sqrt{2}}(\gamma^\mu\gamma^\nu-\gamma^\nu\gamma^\mu)\,.
\end{eqnarray}
Here $\rho=S,\,V,\,T$ and means scalar, vector, and tensor interactions; $\varepsilon,\,\mu=L,\,R$ means left- and right-handed leptons. Each set of indices $\rho$, $\varepsilon$, and $\mu$ uniquely determines the values of $\alpha$ and $\beta$. Ten complex constants $g^\rho_{\varepsilon\mu}$ ($g^T_{RR}=g^T_{LL}\equiv0$) completely describe the weak interaction of charged currents. For the case of the Standard Model, the only nonzero constant is $g^V_{LL}=1$.

It is convenient to express the experimental observables in terms of the Michel parameters, which are bilinear combinations of the coupling constants $g^\rho_{\varepsilon\mu}$. In general, the experimentally observed (integrated over neutrino momenta) differential decay width for $\tlnn$ ($\ell=e$ or $\mu$) can be obtained from the matrix element~\eqref{eq:1}:
\begin{eqnarray}\label{eq:2}
\dfrac{d^2\Gamma}{dx \, d\!\cos{\theta}}=\dfrac{m_\tau}{4\,\pi^3}W_{\ell\tau}^4\,G^2_F\sqrt{x^2-x^2_0}\, \big(F_{IS}(x)\pm F_{AS}(x)P_\tau\cos{\theta} +F_{T_1}(x)P_\tau\sin{\theta}\zeta_1\nonumber\\  + F_{T_2}(x)P_\tau\sin{\theta}\zeta_2 + (\pm F_{IP}(x) + F_{AP}(x)P_\tau\cos{\theta})\zeta_3 \big)\,.
\end{eqnarray}
Here $x=E_\ell/W_{\ell\tau}$ and $x_0=m_\ell/W_{\ell\tau}$ are the reduced energy and mass of the daughter lepton (the ratio of its energy and mass to its maximum energy $W_{\ell\tau}=(m^2_\ell+m^2_\tau)/2m_\tau$, respectively), $\theta$ is the angle between the direction of the daughter lepton momentum and the $\tau$ lepton polarization vector in its rest frame, $\vec{\zeta}=(\zeta_1,~\zeta_2,~\zeta_3)$ is the direction along which the polarization of the daughter lepton is measured. We use the conventional coordinate system $(x_1,~x_2,~x_3)$: the $\vec{x}_3$-axis is directed along with the daughter lepton momentum in the $\tau$ lepton rest system, the $\vec{x}_2 $-axis is directed perpendicular to the $\vec{x}_3$-axis and the $\tau$ lepton polarization $\vec{P}_\tau$, and $\vec{x}_1=\vec{x}_2\times\vec{x}_3$. The plus and minus signs in~\eqref{eq:2} correspond to the $\tau^+$- and $\tau^-$-decays, respectively. The functions $F_{IS}$, $F_{AS}$, $F_{IP}$, $F_{AP}$, $F_{T_1}$, and $F_{T_2}$ are expressed in terms of the Michel parameters and depend only on $x$: 
\begin{eqnarray}\label{eq:3}
F_{IS}(x) & =& x(1-x)+\dfrac{2}{9}\rho(4x^2-3x-x_0^2)+\eta x_0(1-x)\,, \nonumber \\
F_{AS}(x) & = & \dfrac{1}{3}\xi\sqrt{x^2-x_0^2}\left[1-x+\dfrac{2}{3}\delta\left(4x-3-\dfrac{x_0^2}{2}\right)\right], \nonumber \\
F_{IP}(x) & = & \dfrac{1}{54}\sqrt{x^2-x_0^2}\left[-9\xip\left(2x-3+\dfrac{x_0^2}{2}\right)+4\xi\left(\delta-\dfrac{3}{4}\right)\left(4x-3-\dfrac{x_0^2}{2}\right)\right], \\
F_{AP}(x) & = & \dfrac{1}{6}\left[\xipp\left(2x^2-x-x_0^2\right)+4\left(\rho-\dfrac{3}{4}\right)\left(4x^2-3x-x_0^2\right)+2\etapp x_0(1-x)\right],\nonumber \\
F_{T_1}(x) & = & -\dfrac{1}{12}\left[2\left(\xipp+12\left(\rho-\dfrac{3}{4}\right)\right)(1-x)x_0+3\eta(x^2-x_0^2)+\etapp(3x^2-4x+x_0^2)\right], \nonumber \\
F_{T_2}(x) & = & \dfrac{1}{3}\sqrt{x^2-x_0^2}\left(3\dfrac{\alphap}{A}(1-x)+\dfrac{\betap}{A}(2-x_0^2)\right). \nonumber
\end{eqnarray}
Here we have expanded the functions up to the quadratic term $x_0^2$ while holding $\sqrt{x^2-x_0^2}$. Formulas~\eqref{eq:2} and~\eqref{eq:3} are well known and can be found in~\cite{Zyla:2020zbs}. To calculate the partial decay width of $\tlnn$, one should integrate~\eqref{eq:2} over $x$ and $\cos{\theta}$ and sum over the spin of the daughter lepton.

To simplify subsequent calculations, we rewrite formula~\eqref{eq:2}:
\begin{eqnarray}\label{eq:21}
\dfrac{d^2\Gamma}{dx \, d\!\cos{\theta}}=\dfrac{m_\tau}{4\,\pi^3}W_{\ell\tau}^4\,G^2_F\sqrt{x^2-x^2_0}\, \big[G_0 + (\vec{G}\cdot\vec{\zeta})\big]\,,
\end{eqnarray}
where $G_0$ is a scalar and $\vec{G} \equiv (G_1, G_2, G_3)$ is a formal vector formed by functions~(\ref{eq:3}):
\begin{eqnarray}\label{eq:22}
G_0 & = & F_{IS}(x)\pm F_{AS}(x)P_\tau\cos{\theta}\,, \nonumber\\
\vec{G}_{\phantom{0}} & = & \left(F_{T_{1}}(x)P_\tau\sin{\theta}, 
~F_{T_{2}}(x)P_\tau\sin{\theta}, 
~\pm F_{IP}(x) + F_{AP}(x)P_\tau\cos{\theta}\right).
\end{eqnarray}

In the muon decay, $\menn$, the Michel parameters were measured to be consistent with the exact $V-A$ structure with high precision~\cite{Zyla:2020zbs}; therefore, we use the SM muon differential decay width:
\begin{eqnarray}\label{eq:4}
\dfrac{d^2\Gamma}{dy \, d\Omega_e}=\dfrac{G_F^2m_\mu^5}{384\,\pi^4} y^2\left[(3-2y)\pm(2y-1)(\vec{n}_e\cdot\vec{\zeta})\right],
\end{eqnarray}
where $y=2E_e/m_\mu$ is the ratio of the electron energy in the muon rest frame to its maximum value, $\vec{\zeta}$ is the muon polarization direction, $\vec{n}_e$ is the electron momentum direction in the muon rest frame, $d\Omega_e$ is the electron solid angle element. Here plus (minus) sign corresponds to the $\mu^+$ ($\mu^-$) decay. Since we do not register neutrinos in either the $\tau$ lepton decay or the muon decay, the width of the cascade decay $\tau^- \to (\mu^- \to e^-\bar{\nu}_e \nu_\mu) \bar{\nu}_\mu \nu_\tau$ can be obtained by simple muon spin convolution of expressions~\eqref{eq:21} and \eqref{eq:4}. However, the proposed method uses muon decay in a tracking detector, implying muon motion before its decay in a magnetic field, which not only rotates the muon momentum but also drives its spin to precess. This should be taken into account when folding the expressions~\eqref{eq:21} and \eqref{eq:4}. The spin evolution equations were obtained in~\cite{Bargmann:1959gz}. Usually, a constant in time uniform magnetic field is used, which simplifies the spin evolution to the ordinary precession around the axis of the magnetic field. Neglecting the anomalous magnetic moment, the muon spin rotates by the same angle as the momentum: $\phi(t)=\mp2\mu_\mu Ht$, where $t$ is the muon decay time in its rest frame, $H$ is the absolute value of the magnetic field, and $\mu_\mu=e/2m_\mu$ 
is the muon magnetic moment. The upper sign corresponds to a positive charge, and the lower sign corresponds to a negative charge. For the convolution of~\eqref{eq:21} and \eqref{eq:4} over the muon spin, one should replace $\vec{\zeta}$ by $\vec{\zeta}_0$ in \eqref{eq:21} and by $\vec{\zeta}(t)=\boldsymbol{R}(\phi(t))\vec{\zeta_0} $ in \eqref{eq:4}, where the matrix $\boldsymbol{R}(\phi)$ has the form:
\begin{eqnarray}\label{eq:5}
\boldsymbol{R}(\phi)\!=\!\begin{pmatrix}
c + (1 - c) h_1^2 & -h_3 s + (1 - 
        c) h_1 h_2 & h_2 s + (1 - c) h_1 h_3\\
h_3 s + (1 - 
        c) h_1 h_2 &  c + (1 - c) h_2^2 & -h_1 s + (1 - 
        c) h_2 h_3\\
-h_2 s + (1 - 
        c) h_1 h_3 & h_1 s + (1 - c) h_2 h_3 & c + (1 - c) h_3^2
\end{pmatrix}
.
\end{eqnarray}
Here $\vec{h}=(h_1,~h_2,~h_3)$ is the magnetic field direction; $c\equiv\cos{\phi}$ and $s\equiv\sin{\phi}$. The differential width of the cascade decay $\tau^-\to(\mu^-\to e^-\bar{\nu}_e\nu_\mu)\bar{\nu}_\mu\nu_\tau$
is calculated as the spin convolution of the decay width~\eqref{eq:21} and decay width~\eqref{eq:4}:
\begin{eqnarray}\label{eq:6}
\dfrac{d^5\Gamma}{d x\, d\!\cos{\theta}\,d y \,d\Omega_e \,dt} = \mathcal{B}(\mennf)\dfrac{\Gamma_{\tmnnf} }{1-3x_0^2} \, \dfrac{3}{\pi} y^2 \, \sqrt{x^2-x_0^2}\nonumber\\\left[(3-2y)G_0\pm (2y-1)
\vec{n}_e \boldsymbol{R}(\phi) \vec{G}
\right]\dfrac{1}{\tau_\mu} \exp{\left(-\dfrac{t}{\tau_\mu}\right)}\,,
\end{eqnarray}
where $\vec{n}_e$ is the electron momentum direction in the muon rest frame, written in the coordinate system $(x_1,~x_2,~x_3)$, $\Gamma_{\tmnnf}$ is the partial width of the $\tmnn$ decay, $\mathcal{B}(\mennf)$ is the branching ratio of the $\menn$ decay, $\tau_\mu$ is the muon lifetime. The plus sign and the minus sign correspond to the $\tau^+$- and $\tau^-$-decays, respectively.

Expression~\eqref{eq:6} is applicable for \ee\ annihilation at any $\sqrt{s}$ with the proviso that we know the momentum of the $\tau$ lepton and its polarization. Hereinafter, we assume that the momentum of the $\tau$ lepton is known. In real experiment, due to the presence of undetected neutrinos, it is necessary to take into account the uncertainty in the $\tau$ momentum by integration over the region of possible $\tau$ directions.

Thus, we have obtained the differential width of the cascade $\tau^-\to(\mu^-\to e^-\bar{\nu}_e\nu_\mu)\bar{\nu}_\mu\nu_\tau$ decay, taking into account the muon rotation in the detector magnetic field. Further analysis of the obtained expression is carried out in the context of its application in the conditions of a particular experiment. It is worth noting that due to the rarity of muon decays in the detector, the measurement accuracy of the Michel parameters \xip, \xipp, \etapp, $\alphap/A$, and $\betap/A$ will be significantly lower than those of $\rho$, $\eta$, $\xi$, and $\xi\delta$ within the same experiment; therefore, to simplify all further calculations, the latter can be equated to their SM values.

\section{Application in the experiments at \texorpdfstring{$\ee$}{TEXT} colliders \label{sec:app}}

For the experimental implementation of the proposed method, three conditions are necessary: a large high-purity $\tau$ sample, determination of the $\tau$ momentum, and, finally, knowledge of the $\tau$ polarization before its decay. The latter can be abandoned for the measurement of \xip, which is not related to the mother lepton polarization. All the conditions are fully met at the future Super Charm-Tau Factory (SCTF) with polarized beam and partially met at Belle II experiment, where tagging with the second $\tau$ in the event helps to constrain the directions of the signal $\tau$ momentum and polarization.

\subsection{SCTF with polarized beam\label{subsec:ctau}}

The SCTF experiment with a polarized beam provides an ideal environment for the simplest and the most accurate measurement of the Michel parameters, which describe daughter muon polarization. In this experiment, $\tau$ leptons will be produced almost at rest and with a known polarization. Moreover, the $\ee\to \tat$ events can be effectively selected with a high purity of the \tat\ sample.

\begin{figure}[tbp]
\centering
  \includegraphics[width=0.55\linewidth]{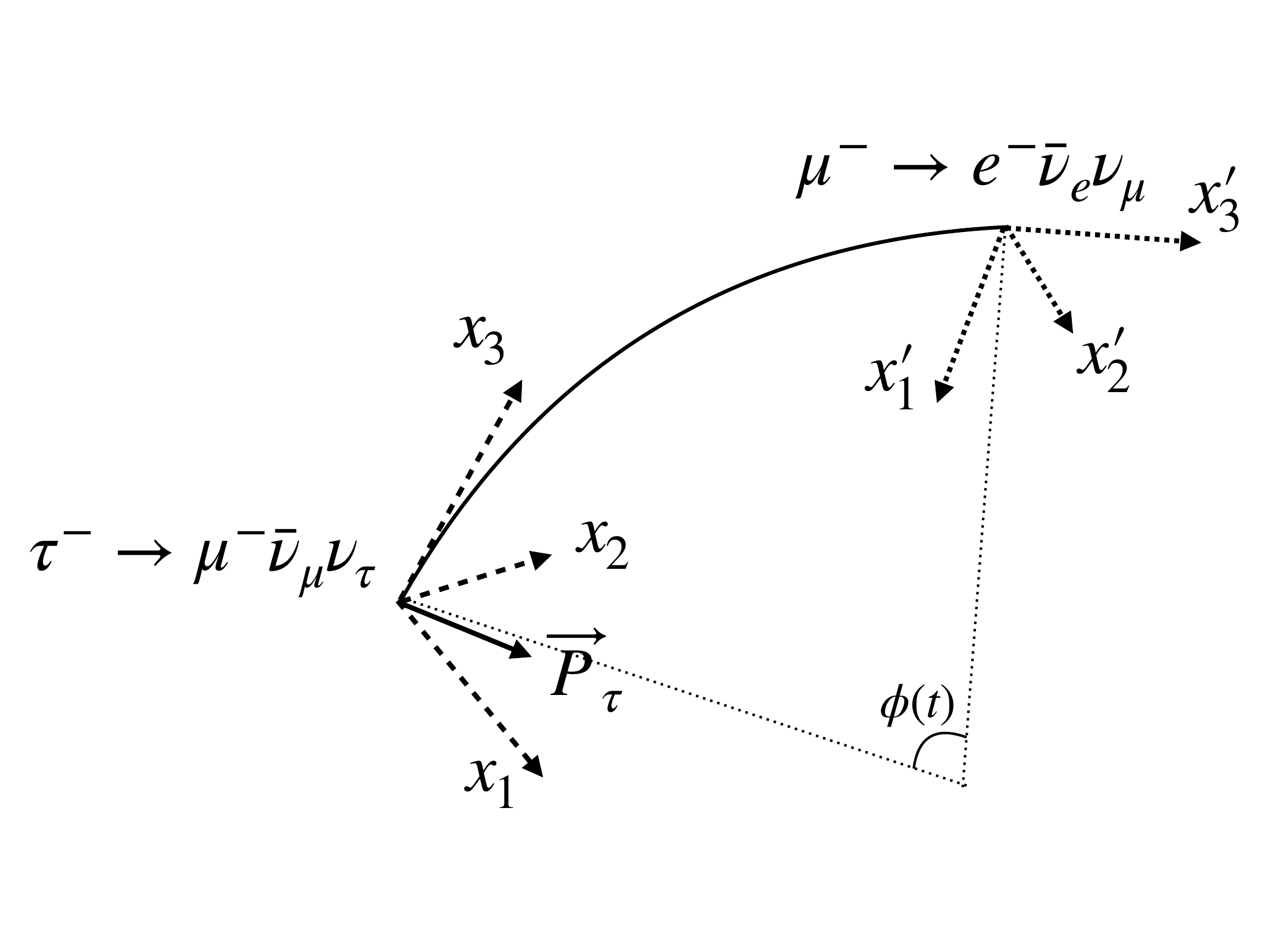}
\caption{$(x_1,~x_2,~x_3)$ and $(x_1',~x_2',~x_3')$ coordinates illustration for the SCTF case of the cascade decay.}
\label{fig:x1x2x3_SCTF}
\end{figure}

The expression for the differential width of the cascade decay~\eqref{eq:6} can be simplified by a transition to a new coordinate system $(x_1',~x_2',~x_3')$ defined at the muon decay vertex. This coordinate system is obtained by rotation of the initial coordinate system $(x_1,~x_2,~x_3)$ by the angle $\phi$ around the magnetic field vector, as shown in figure~\ref{fig:x1x2x3_SCTF}.
This transformation leads to $\vec{n}_e\to\vec{n}_e'$ and $y^2\, dy\, d\Omega_e\, dt \to |J| y'{}^2\, dy'\, d\Omega_e'\, dt'$, where $t'=t$ and $y'=y$. The latter is due to the fact that our transformation is a rotation. For the same reason, the Jacobian determinant, $|J|$, is equal to unity.\footnote{The simplest way to prove it is to move from spherical coordinates $(y, \,\Omega)$ to Cartesian coordinates $(n_1,\,n_2,\,n_3)$:
\begin{eqnarray}\label{eq:71}
|J|\!=\!\left| \dfrac{\partial(n_1',\,n_2',\,n_3',\,t')}{\partial(n_1,\,n_2,\,n_3,\,t)}\right|=
\begin{vmatrix}
\boldsymbol{R}^T_{ij} & \dfrac{d\boldsymbol{R}^T_{ij}}{dt} n_j \\
0 & 1 
\end{vmatrix}=\left| \boldsymbol{R}^T_{ij} \right|=1\,.\nonumber
\end{eqnarray}}
After the transformation, the differential decay width takes the following form:
\begin{eqnarray}\label{eq:72}
\dfrac{d^5\Gamma}{d x \, d\!\cos{\theta}\,dy \,d\Omega_e'\,dt} = \mathcal{B}(\mennf)\dfrac{\Gamma_{\tmnnf}}{1-3x_0^2}\dfrac{3}{\pi} y^2\sqrt{x^2-x_0^2}\nonumber\\\left[(3-2y)G_0\pm (2y-1)(\vec{n}_e' \cdot \vec{G}) \right] \dfrac{1}{\tau_\mu}\exp{\left(-\dfrac{t}{\tau_\mu}\right)}\,. 
\end{eqnarray}
After we present the electron momentum in the new coordinate system for each decay event, the time dependence disappears from angular distributions and remains only in the exponential. 
Finally, integrating over $t$ and substituting $\vec{n}_e'=(\sin{\theta_e}\cos{\psi_e},\,\sin{\theta_e}\sin{\psi_e},\,\cos{\theta_e})$, we can derive the explicit dependence of the differential decay width on $\cos{\theta_e}$ and $\psi_e$:
\begin{eqnarray}\label{eq:7}
\dfrac{d^5\Gamma}{dx\, d\!\cos{\theta}\, dy\, d\!\cos{\theta_e}\, d\psi_e} = \mathcal{B}(\mennf)\dfrac{\Gamma_{\tmnnf}}{1-3x_0^2}\dfrac{3}{\pi} y^2\sqrt{x^2-x_0^2} \left[(3-2y)G_0 \right.\nonumber\\\left. \pm (2y-1)(G_1\sin{\theta_e}\cos{\psi_e} + G_2\sin{\theta_e}\sin{\psi_e} + G_3\cos{\theta_e}) \right]. 
\end{eqnarray}
The muon longitudinal polarization, determined by the parameters $\xip$ and $\xipp$, is frozen (up to $g_\mu-2$ corrections) into the $x_3'$-axis, thus, directed along with the muon momentum in the $\tau$ rest frame (which is well approximated by the laboratory frame at SCTF). Finally, after the integration of~\eqref{eq:7} over $\psi_e$, only the dependence on the muon longitudinal polarization remains:
\begin{eqnarray}\label{eq:74}
\dfrac{d^4\Gamma}{dx\, d\!\cos{\theta}\, dy\, d\!\cos{\theta_e}} = \mathcal{B}(\mennf)\dfrac{6\Gamma_{\tmnnf}}{1-3x_0^2}\,y^2\sqrt{x^2-x_0^2}\nonumber\\\left[(3-2y)G_0\pm (2y-1) G_3\cos{\theta_e} \right].
\end{eqnarray}

Using the obtained formulas, we estimate the SCTF sensitivity to the Michel parameters $\xip$, $\xipp$, $\etapp$, $\alphap/A$, and $\betap/A$ measurements using the SCTF parameters from~\cite{Bondar:2013cja}. In this project, it is planned to use a highly polarized electron beam (the degree of polarization of the beam at the interaction point is $\xi_\text{beam}=0.8$) and an unpolarized positron beam. For the SCTF energies, as shown in~\cite{Tsai:1994rc}, the degree of $\tau$ lepton polarization and its direction with good accuracy are independent of the $\tau$ lepton production angle. We thus fix the degree of the $\tau$ polarization to be $P_\tau\approx\xi_\text{beam}=0.8$ and its direction to be along the electron beam. To simplify the estimates, we assume that full SCTF statistics will be collected around the $\tau^+\tau^-$-pair production threshold; thus, $\tau$ leptons are produced almost at rest.

Since there is no full detector simulation and reconstruction program, we use a toy Monte Carlo (MC) simulation with an overall efficiency to reconstruct the decay of interest defined as $\eta=\eta_\text{tag}\,\omega_\text{dec}\,\eta_\text{kink}\eta_\text{sel}$. Here $\eta_\text{tag}\approx 30\%$ is the selection efficiency of the $\tau^+\tau^-$ events based on the estimation for the BES~III experiment~\cite{Asner:2008nq}. We consider only the simplest cases for the reconstruction algorithm to find the kink: the muon decays in the drift chamber (DC) on the first turn of its track and in $\geq\!10\,\text{cm}$ from the outer walls. These requirements ensure the reconstruction of the kink by the track reconstruction algorithm with a typical efficiency $\eta_\text{kink}\approx 90\%$ for both mother and daughter tracks. Using MC simulation, we calculate the probability for muon to decay with these requirements to be $\omega_\text{dec}\approx 3.2\cdot10^{-4}$. 

Our requirement for the decay vertex in the DC provides at least 10 hits for the secondary electron; thus, its momentum and angle resolution will be quite good. It was estimated in~\cite{Bodrov:2021hfe} given the spatial hit resolution in the DC of about $125\,\mu\text{m}$~\cite{Bondar:2013cja}. Here we have confirmed that such a resolution does not affect the accuracy of the Michel parameters measurement.

The last factor that reduces the analyzed sample, $\eta_\text{sel}$, is related to the need to suppress backgrounds that initially exceed the signal.
The expected background contamination from other processes rather than $\tau^+ \tau^-$, mainly $q\bar{q}$, was estimated in~\cite{Asner:2008nq} to be $\sim6\%$, and we ignore it. The main background sources come from $\tau^+\tau^-$-pairs events with kink candidates that imitate the signal process. They are charged pion and kaon decays, and the particles elastic scattering. The specific kinematics allow effectively discriminate background processes from signal one: pion and kaon decay mainly to two monochromatic particles producing a narrow line in the mother particle rest frame, while elastic scattering conserves the momentum magnitude of the particle. The efficiency of the discrimination is determined by the momentum resolution, which is good enough; thus, we estimate that the backgrounds can be suppressed to a negligible level with $\eta_\text{sel}\approx 80\%$ efficiency for the signal~\cite{Bodrov:2021hfe}. 

\begin{figure}[tbp]
\centering
  \includegraphics[width=0.72\linewidth]{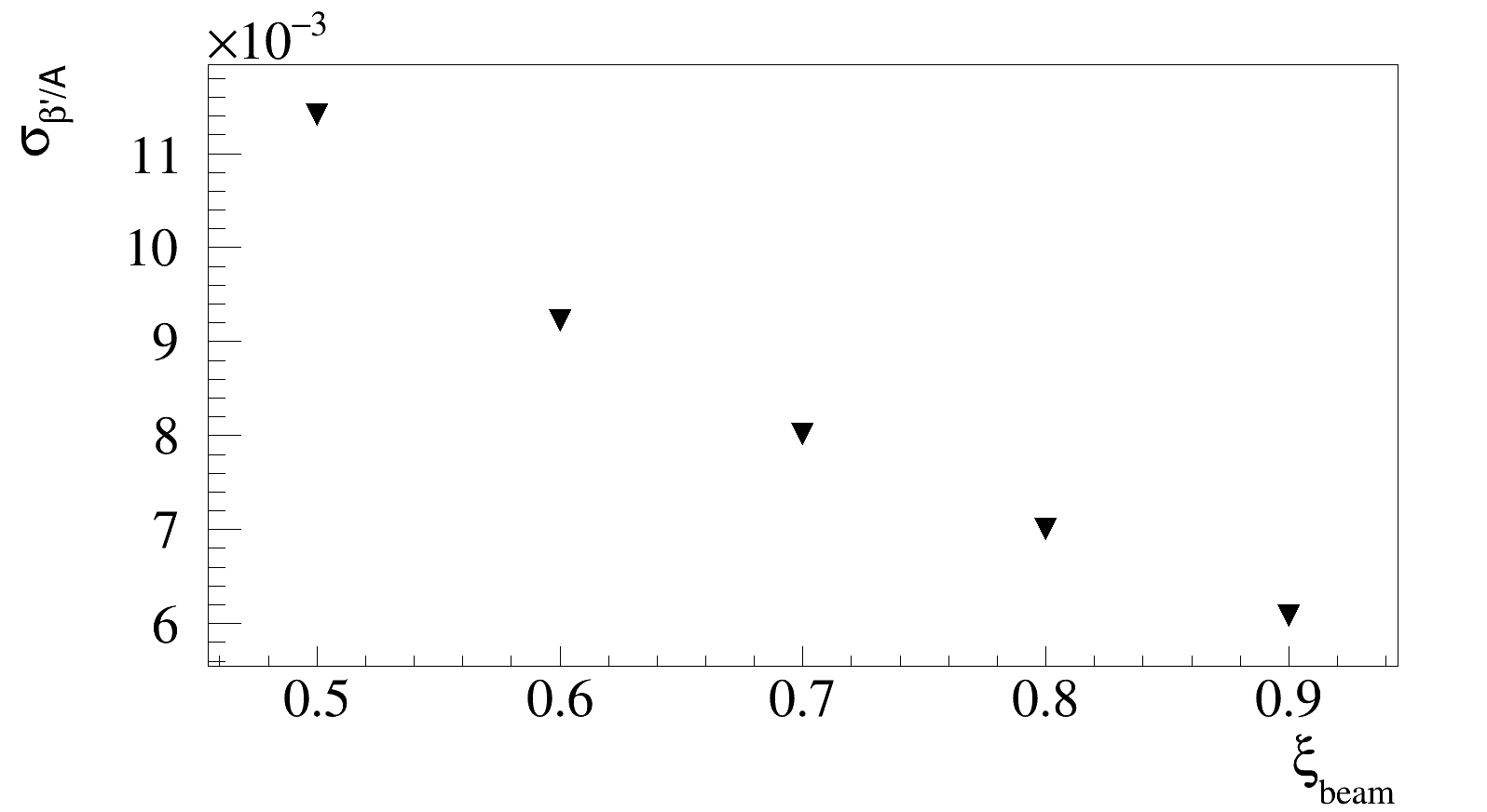}
\caption{The dependence of the expected uncertainty of the Michel parameter $\betap/A$ measurement on the beam polarization.}
\label{fig:error_vs_pol}
\end{figure}

Taking into account the branching fraction of the $\tmnn$ decay, we obtain the number of signal kink events for the $\tau$ decay sample, expected with the full SCTF statistics, to be $N\approx5\cdot10^5$. Using the generated toy MC sample with the statistics corresponding to the full data set at SCTF, we estimate the sensitivity of a global fit to the SCTF data that includes all Michel parameters. We separately analyze $\tau^+$ and $\tau^-$ samples and, for each of them, estimate the precision of the Michel parameters measurement. We perform a 5D unbinned likelihood fit of the simulated data on the $x$, $\cos{\theta}$, $y$, $\cos{\theta_e}$, and $\psi_e$ variables with the fit function given by~\eqref{eq:7}. The free parameters of the fit are only those Michel parameters that determine the muon polarization ($\xip$, $\xipp$, $\etapp$, $\alphap/A$, and $\betap/A$), while parameters $\rho$, $\eta$, $\xi$, and $\xi\delta$ are fixed to their SM values since they can be measured separately with unprecedented statistical accuracy at SCTF. We obtained the following statistical errors from the fit:
$\sigma_{\xip}\approx6\cdot10^{-3}$, $\sigma_{\xipp}\approx3\cdot10^{-2}$, $\sigma_{\etapp}\approx2\cdot10^{-2}$, $\sigma_{\alphap/A}\approx14\cdot10^{-3}$, and $\sigma_{\betap/A}\approx7\cdot10^{-3}$. 
The expected SCTF accuracy is comparable to those achieved in the muon decay.
We also present the dependence of the expected uncertainty of the Michel parameter $\betap/A$ measurement on the polarization of the beam (figure~\ref{fig:error_vs_pol}). The uncertainty of other Michel parameters scales with the beam polarization in the same way.

In case of a discrepancy with the SM prediction is observed, it would be possible to find the Lorentz structure of new interactions. Otherwise, global fit allows setting upper limits on the magnitude of the coupling constants $g^\rho_{\varepsilon\mu}$. They are presented in table~\ref{tab:coup_const_PDG} together with the current loose limits~\cite{Zyla:2020zbs} for comparison. 

We estimate the potential limits using the accuracy for the Michel parameters $\xip$, $\xipp$, $\etapp$, $\alphap/A$, and $\betap/A$ obtained in this work. For the Michel parameters $\rho$, $\eta$, $\xi$, and $\xi\delta$, it was shown in work~\cite{Epifanov:2020elk} that the statistical precision will be at the $10^{-4}$ level. Actually, the Belle experiment has already achieved $10^{-3}$ statistical uncertainty. However, the systematic uncertainties dominate this measurement at the percent level~\cite{Epifanov:2017kly}. That is why we set $10^{-3}$ uncertainty for $\rho$, $\eta$, $\xi$, and $\xi\delta$, assuming that systematic will be controlled at that level.

To set limits on the coupling constants magnitude $|g^\rho_{\varepsilon\mu}|$, we use the alternative parametrization from~\cite{Fetscher:1986uj}: $Q_{RR}$, $Q_{LR}$, $Q_{RL}$, $Q_{LL}$, $B_{LR}$, and $B_{RL}$. The Michel parameters $\rho$, $\xi$, $\xi\delta$, $\xip$, and $\xipp$ expressions in terms of these variables can be found in~\cite{Gagliardi:2005fg}. For the sake of simplicity, we do not follow the full procedure of finding boundaries and do not use information about $\eta$, $\etapp$, $\alphap/A$, and $\betap/A$. Instead, we refit our simulated data sample with new variables with constraints on $\rho$, $\xi$, and $\xi\delta$ to extract the uncertainties, which are then utilized to set limits. The values of variables are set to their SM expectations. The result for $|g^\rho_{\varepsilon\mu}|$ constraints, obtained for the $\tau^-$-decay sample, is shown in table~\ref{tab:coup_const_PDG}. We do not show results for $|g^S_{LL}|$ and $|g^V_{LL}|$ because they cannot be obtained separately in $\tau$-decays measurements. The same result is for the $\tau^+$-decay sample. We analyze the samples for both charges separately as it will allow testing $CPT$-invariance, while for the muon decay, there is no such an opportunity because not all Michel parameters were measured for both signs. 
\begin{table}[tbp]
\caption {Here coupling constants $g^\rho_{\varepsilon\mu}$ are presented with 95\% confidence level experimental limits and 90\% confidence level limits from our estimations. Our results are shown only for the $\tmnn$ sample, while for the current experimental limits, results for both signs were averaged.} 
\label{tab:coup_const_PDG} 
\begin{center}
\begin{tabular}{l l l | l l l | l l l}
 \hline
 \hline
& ref.~\cite{Zyla:2020zbs} & SCTF & & ref.~\cite{Zyla:2020zbs}  & SCTF & & ref.~\cite{Zyla:2020zbs}  & SCTF \\ \hline \hline
 $|g^S_{RR}|$ & $< 0.72$ & $< 0.18$ & $|g^V_{RR}|$ & $ < 0.18$ & $ < 0.09$ & $|g^T_{RR}|$ & $\equiv 0$ & $\equiv 0$\\
 $|g^S_{LR}|$ & $< 0.95$ & $< 0.18$ & $|g^V_{LR}|$ & $ < 0.12$ & $ < 0.05$ & $|g^T_{LR}|$ & $ < 0.079$ & $< 0.03$\\
 $|g^S_{RL}|$ & $< 2.01$ & $< 0.19$ & $|g^V_{RL}|$ & $ < 0.52$ & $ < 0.05$ & $|g^T_{RL}|$ & $ < 0.51$ & $< 0.03$\\
 $|g^S_{LL}|$ & $< 2.01$ &  & $|g^V_{LL}|$ & $ < 1.005$ &  & $|g^T_{LL}|$ & $\equiv 0$ & $\equiv 0$\\
 \hline
 \hline
\end{tabular}
\end{center}
\end{table}

\begin{figure}[tbp]
\centering
  \includegraphics[width=0.72\linewidth]{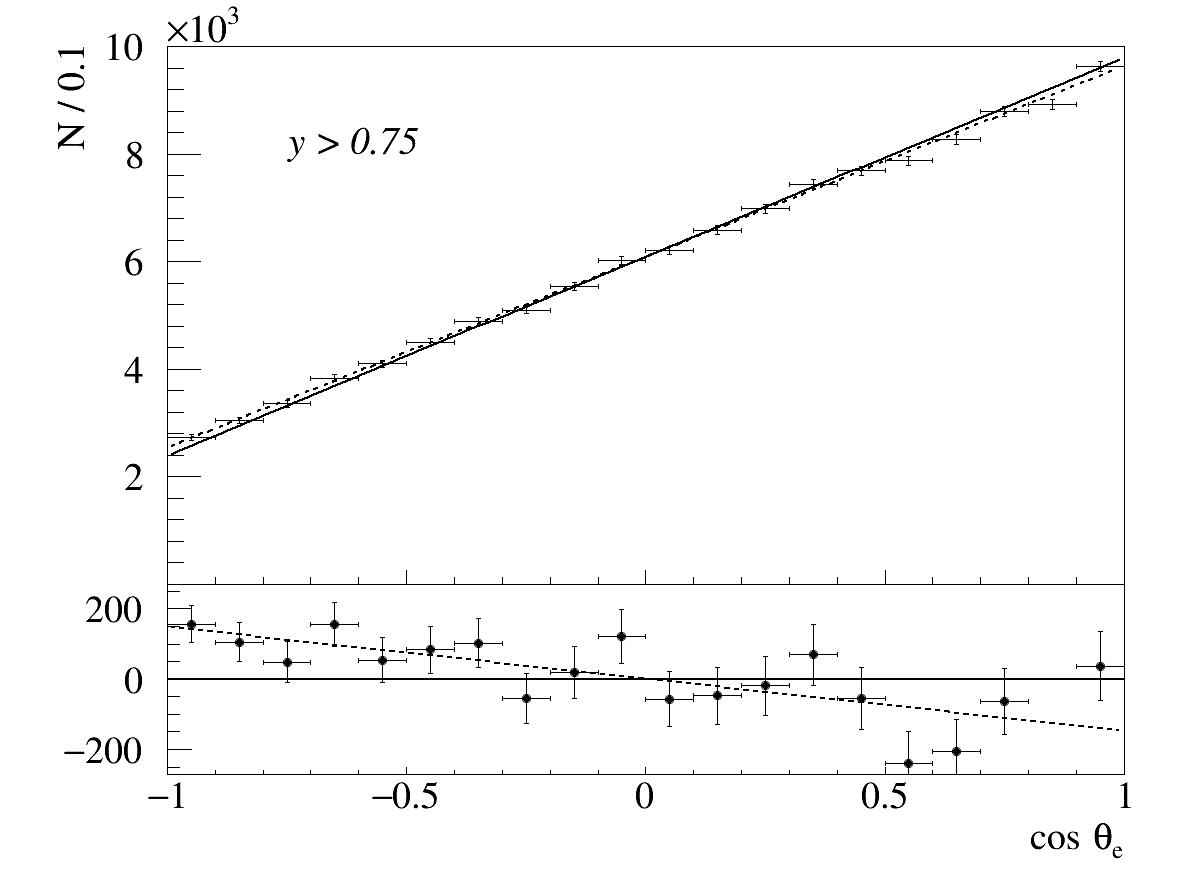}
\caption{The projection on $\cos{\theta_e}$ for $y>0.75$. Points with error bars are the simulated data with $\xip=0.96$ at SCTF, the solid line is the SM expectation, the dashed line is the fit function. The bottom plot shows the subtraction of the solid line from the data points.}
\label{fig:MP_xip_SCTF}
\end{figure}

We demonstrate the ability of SCTF using the proposed method to discover new physics if it biases the Michel parameters. To illustrate this, we generated Monte Carlo samples with a small contribution of the NP with a different Lorentz structure to the $\tmnn$ decay. In the first sample, we slightly shifted from the SM value the Michel parameter $\xip$ ($\xipp=\xip=0.96$) that determines the muon longitudinal polarization ($P_L$) and the probability of any (right- or left-handed) $\tau$ lepton to decay to the right-handed muon: $Q^\mu_{R}=(1-\xip)/2$. Such a case can be implemented in the NP scenario with an additional admixture of the scalar interaction that does not change other Michel parameters. The points with error bars in figure~\ref{fig:MP_xip_SCTF} represent a projection onto $\cos{\theta_e}$ for the $y>0.75$ interval\footnote{Integration of expression~\eqref{eq:7} over $y$ in the whole interval partially cancels the angular dependence of the muon polarization term; thus, we choose the interval where the effect is maximum.} for the simulated events with $\xip=0.96$ corresponding to the full SCTF data sample. The solid line corresponds to the SM value of $\xip=1$, and the dashed line shows the fit function. Although the shown projection has reduced sensitivity compared to 5D-fit, the difference between the data and the SM expectation is clearly visible. The statistical significance of the deviation from the SM in the fit is greater than $5\sigma$, pointing out that the NP leading to $\xip<0.96$ can be revealed at SCTF.

\begin{figure}[tbp]
\centering
  \includegraphics[width=0.72\linewidth]{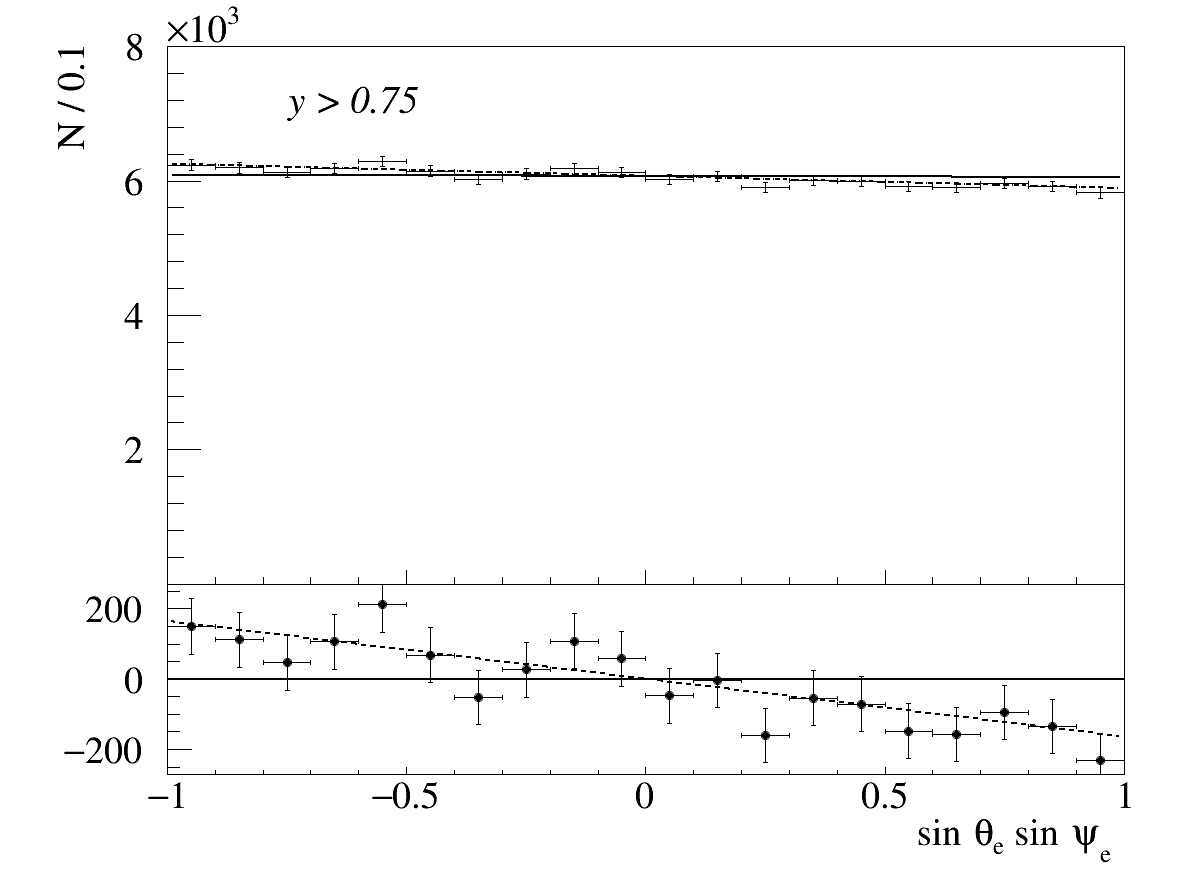}
\caption{The projection on $\sin{\theta_e}\sin{\psi_e}$ for $y>0.75$. Points with error bars are the simulated data at SCTF with $\alphap/A=0$ and $\betap/A=0.03$, the solid line shows SM expectation, the dashed line is a fit function. The bottom plot shows the subtraction of the solid line from the data points.}
\label{fig:MP_betap_SCTF}
\end{figure}

The second sample is dedicated to another NP scenario that affects the muon transverse polarization, $P_{T_2}$. It is determined by two Michel parameters, $\alphap/A$ and $\betap/A$. Non-identical equality of the transverse polarization, $P_{T_2}$, to 0 (and thus $\alphap/A \neq 0$ and $\betap/A \neq 0$) means the $T$-violating NP contribution. In the simulation, we set $\alphap/A=0$ and $\betap/A=0.03$. The projection onto $\sin{\theta_e}\sin{\psi_e}$ for $y>0.75$ for this sample is shown in figure~\ref{fig:MP_betap_SCTF}. 
Points with error bars show the simulated data corresponding to the full SCTF data set. The solid line corresponds to the SM case ($\alphap/A=0$ and $\betap/A=0$), and the dashed line shows the fit function. The fit yields $5\sigma$ significance of the NP contribution; thus, SCTF is able to observe the $T$-violating NP effect if it results in $\betap/A\gtrsim 0.03$.

\subsection{Belle~II experiment\label{subsec:belle2}}

From expression~\eqref{eq:6}, we can conclude that to measure the Michel parameter $\xip$, it is not necessary to know the $\tau$ lepton polarization, while the parameters $\xipp$, $\etapp$, $\alphap/A$, and $\betap/A$ cannot be measured without this information. At the Belle~II experiment, $\tau$ leptons are produced with zero average polarization. However, it is possible to extract information about the polarization of the signal $\tau$ using the decay of the second (tagging) $\tau$ lepton in the event. This procedure complicates the calculations of the theoretical function, and experimental analysis becomes difficult due to the higher dimensional fit function. Moreover, the accuracy of such measurement will be lower compared to the SCTF experiment since not all decays of the tagging $\tau$ are useful, and besides, the information on the $\tau$ spin is indirect (statistical). We thus consider the application of the method to measure only the Michel parameter \xip\ and show that its measurement has comparable to the SCTF accuracy.

\begin{figure}[tbp]
\centering
  \includegraphics[width=0.55\linewidth]{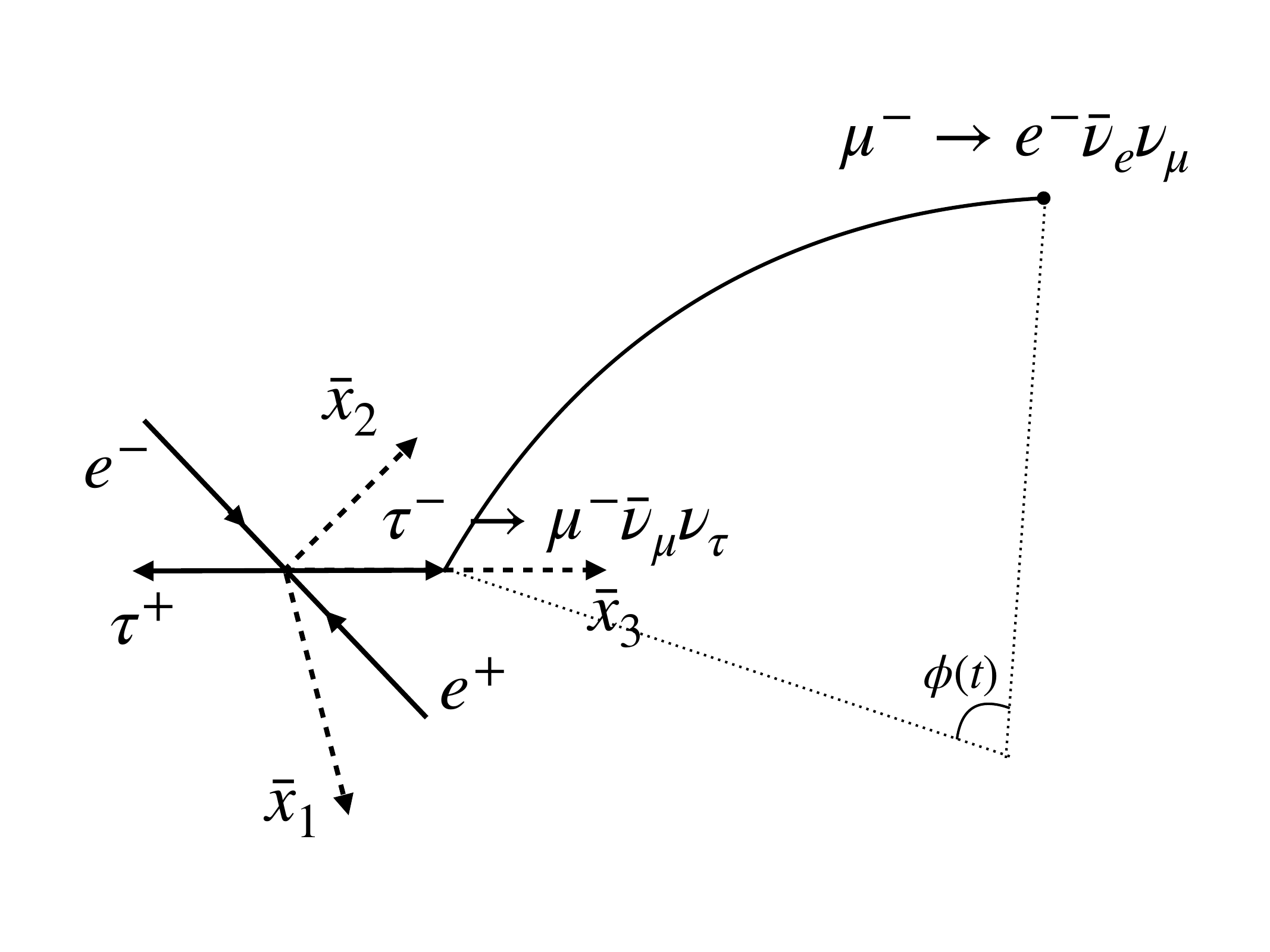}
\caption{$(\bar{x}_1,~\bar{x}_2,~\bar{x}_3)$ coordinates illustration for the Belle~II case of the cascade decay.}
\label{fig:x1x2x3_Belle2}
\end{figure}

For $\xip$ measurement at Belle II case, we first average differential decay width over $\tau$ polarization. For this, one cannot use expression~\eqref{eq:6} because it is written in the coordinate system given by the $\tau$ polarization vector; instead, one should use the conventional for the studies of the $\ee\to\tat$ process with unpolarized beams coordinate system. This system, $(\bar{x}_1,~\bar{x}_2,~\bar{x}_3)$, is defined as follows: the $\bar{x}_3$-axis is directed along with the $\tau^-$ momentum in the CMS, the $\bar{x}_2$-axis is perpendicular to the $\bar{x}_3$-axis and the direction of the electron beam in the CMS, and the $\bar{x}_1$-axis forms with the $\bar{x}_2$- and $\bar{x}_3$-axis right-handed orientation. Figure~\ref{fig:x1x2x3_Belle2} illustrates the bar coordinate system with a schematic view of the cascade decay we study.  

The differential decay width of the $\tmnn$ decay written in the coordinate system $(\bar{x}_1,~\bar{x}_2,~\bar{x}_3)$ has the following form:
\begin{eqnarray}\label{eq:91}
\dfrac{d^2\Gamma}{dx\, d\Omega_\mu}=\dfrac{\Gamma_{\tmnnf}}{1-3x_0^2}\dfrac{3}{2\pi}\sqrt{x^2-x_0^2}\left[F_{IS}(x) \pm F_{IP}(x)\, (\vec{n}_\mu \cdot\vec{\zeta})\right.\nonumber\\ \left. \pm F_{AS}(x)\, (\vec{n}_\mu \cdot \vec{s}) + (\vec{B} \cdot \vec{s})\right],
\end{eqnarray}
where
\begin{eqnarray}\label{eq:92}
\vec{B} = F_{T_1}(x)\,\vec{\zeta}+[F_{AP}(x)-F_{T_1}(x)]\,(\vec{\zeta}\cdot\vec{n}_\mu)\,\vec{n}_\mu + F_{T_2}(x)\,[\vec{\zeta}\times\vec{n}_\mu]\,.
\end{eqnarray}
Here $\vec{s}$ is the $\tau$ spin direction, and $\vec{n}_\mu$ is the direction of muon emission in the $\tau$ rest frame, $d\Omega_\mu$ is the muon solid angle element, and $\vec{\zeta}$ is the muon spin direction in its rest frame. After averaging over $\tau$ lepton spin, expression~\eqref{eq:91} simplifies to the following:
\begin{eqnarray}\label{eq:93}
\dfrac{d^2\Gamma}{dx\, d\Omega_\mu}=\dfrac{\Gamma_{\tmnnf}}{1-3x_0^2}\dfrac{3}{2\pi}\sqrt{x^2-x_0^2}\left[F_{IS}(x) \pm  F_{IP}(x)\, (\vec{n}_\mu \cdot\vec{\zeta}) \right].
\end{eqnarray}
Here $F_{IP}(x)$ contains the Michel parameter $\xip$ and describes the muon longitudinal polarization. 

To obtain the differential decay width of the cascade decay, we should repeat the folding procedure of the $\tmnn$ differential decay width~\eqref{eq:93} with the $\menn$ differential decay width~\eqref{eq:4} described in section~\ref{sec:theory}. This leads to the following result:
\begin{eqnarray}\label{eq:8}
\dfrac{d^5\Gamma}{d x\,d\Omega_\mu\,d y \,d\Omega_e \,dt} = \mathcal{B}(\mennf)\dfrac{\Gamma_{\tmnnf}}{1-3x_0^2}\dfrac{3}{2\pi^2} y^2\sqrt{x^2-x_0^2}\left[(3-2y)F_{IS}(x) \right.\nonumber \\ \left.+ (2y-1)F_{IP}(x)
\vec{n}_e \boldsymbol{R}(\phi) \vec{n}_\mu
\right]\dfrac{1}{\tau_\mu} \exp{\left(-\dfrac{t}{\tau_\mu}\right)}\,.
\end{eqnarray}

We then repeat the procedure of the coordinate system rotation to the vertex of the muon decay after propagation in the magnetic field, explained in the previous section, and integrate over time:
\begin{eqnarray}\label{eq:81}
\dfrac{d^4\Gamma}{d x\,d\Omega_\mu\,d y \,d\Omega_e'} = \mathcal{B}(\mennf)\dfrac{\Gamma_{\tmnnf}}{1-3x_0^2}\dfrac{3}{2\pi^2} y^2\sqrt{x^2-x_0^2}\nonumber\\\left[(3-2y)F_{IS}(x)+ (2y-1)F_{IP}(x)
(\vec{n}_e'\cdot \vec{n}_\mu)
\right],
\end{eqnarray}
which can be simplified to 
\begin{eqnarray}\label{eq:82}
\dfrac{d^3\Gamma}{d x\,d y \,d\cos{\theta_e'}} = \mathcal{B}(\mennf)\dfrac{12\Gamma_{\tmnnf}}{1-3x_0^2} y^2\sqrt{x^2-x_0^2}\nonumber\\\left[(3-2y)F_{IS}(x)+ (2y-1)F_{IP}(x)
\cos{\theta_e'}\right].
\end{eqnarray}
Here we have introduced $\theta'_e$, the angle between $\vec{n}_e'$ and $\vec{n}_{\mu}$. As it was noted before, since the direction of the $\tau$ lepton cannot be precisely determined due to the undetectable neutrino, additional averaging over the kinematically allowed $\tau$ direction is required. This procedure is commonly used in this kind of analysis.

Using the obtained theoretical expression, we estimate a sensitivity for the Michel parameter $\xip$ measurement at the Belle~II experiment. Again, we use a toy Monte Carlo simulation with an overall efficiency to reconstruct the decay of interest defined as $\eta=\eta_\text{tag}\,\omega_\text{dec}\,\eta_\text{kink}\,\eta_\text{sel}$. Here $\eta_\text{tag}$ is the selection efficiency of the $\tau^+\tau^-$ events at B-factories; its typical value at Belle is equal to 15\%~\cite{Epifanov:2017kly}. 

The probability of the muon to decay inside CDC ($\geq10\,\text{cm}$ from the walls) and on the first turn is calculated using the MC simulation to be $\omega_\text{dec}\approx 2.8\cdot10^{-4}$. While the outer radius of the drift chamber is larger at Belle~II, the muon boost is also larger; thus, in total, the probability to decay within kink reconstructable volume is slightly smaller than at SCTF. Since the kink reconstruction and selection algorithms are the same as for the SCTF, the efficiencies $\eta_\text{kink}\approx90\%$ and $\eta_\text{sel}\approx80\%$ are also supposed to be the same.

Finally, the number of reconstructed cascades $\tau^-\to (\menn)\bar{\nu}_\mu\nu_\tau$ from the full expected Belle~II data sample is $N \approx 4.8\cdot10^5$. We estimate the accuracy of the Michel parameter $\xip$ measurement in the Belle~II experiment to be $\sigma_{\xip}\approx7\cdot10^{-3}$ for the separate analysis of the $\tau^+$ and $\tau^-$ decay samples. This result is comparable to one obtained for the SCTF.

\section{Systematic uncertainties\label{sec:syst}}

As shown above, the expected statistical accuracy in all Michel parameters in $\tau$-decays will be comparable to those achieved in the $\menn$ decay. However, it is necessary to control the systematic uncertainties at the same level. In this paper, we do not consider in detail this problem since its solution depends on the specific implementation of the experiment. We just briefly discuss the expected main sources of systematic uncertainties and how to evaluate them. 

The major sources of the systematic errors are the uncertainties in the efficiency of the signal process reconstruction (depending on the kinematics of muon decay) and the remaining background calculation.
Concerning the first one, the efficiency of the $\menn$ kink reconstruction strongly depends on the direction of the daughter electron emission in the muon rest frame. This indeed needs to be precisely known for which one can use the background processes. Inverting the veto allows selecting huge samples of the $\pi^-\to\mu^-\bar{\nu}_\mu$ or $K^-\to\mu^-\bar{\nu}_\mu$ decays. As pseudoscalars decays are uniform without any model uncertainties, one can obtain the efficiency directly from the data. Moreover, the statistics of hadronic kinks exceed the signal statistics; thus, we can conclude that the systematics due to detector response uncertainties is smaller than statistical error.

The background can be extracted using the MC simulation, as all background processes are well known and reliably described in the simulation. Indeed, the kinematics of $\ee\to\tat$ process and $\tau$ lepton decays are well reproduced by the KKMC~\cite{Jadach:1999vf} and TAUOLA~\cite{Was:2000st, Jadach:1993hs} Monte Carlo generators. This is confirmed experimentally by using these generators for much more precise measurements of other Michel parameters. The estimation of the detector effects is tied to a specific experiment. Nevertheless, there are always statistically large control samples of tagged pion and kaon kinks, for example, from $D$-decays, that can be used to control the background calculation.

Thus, we conclude that systematic errors can be controlled at least at the same level as statistical ones using the data.

\section{Conclusion\label{sec:conc}}

In this work, the method of the first direct measurement of all Michel parameters, which determine the polarization of the daughter muon from the $\tmnn$ decay, was described in detail. In addition, the feasibility study of this method application at the ongoing experiment Belle~II and the future Super Charm-Tau Factory was carried out. For the first one, only the Michel parameter $\xi'$ measurement was discussed, while for the SCTF, parameters $\xi'$, $\xi''$, $\eta''$, $\alpha'/A$, and $\beta'/A$ were considered. The potential accuracy of the Michel parameters measurement was estimated to be comparable to those achieved in the muon decay. It was shown that the SCTF with polarized beam is an optimal experiment for the precise measurement of all Michel parameters in $\tau$-decays.

\section{Acknowledgments\label{sec:acknow}}
The authors would like to thank D.~Epifanov for critical remarks and useful discussions. The work of D.~Bodrov was carried out within the framework of the Basic Research Program at the National Research University Higher School of Economics (HSE). P.~Pakhlov acknowledges the support by the Russian Science Foundation under contract 22-22-00564.

\bibliography{bibl}

\end{document}